# Super-efficiency and Stock Market Valuation:

Evidence from Listed Banks in China (2006 to 2023)

Yun Liao [1]

---

[1] Author affiliation: Credit Management School, Guangdong University of Finance, Guangzhou, China. Address: No. 527, Yingfu Road, Tianhe District, Guangzhou City, China. Email: 47-066@gduf.edu.cn

**Super-efficiency and Stock Market Valuation:**

Evidence from Listed Banks in China (2006 to 2023)

Abstract: This study investigates the relationship between bank efficiency and stock market valuation using an unbalanced panel dataset of 42 listed banks in China from 2006 to 2023. We employ a non-radial and non-oriented slack based super-efficiency Data Envelopment Analysis (Super-SBM-UND-VRS based DEA) model, which treats Non-Performing Loans (NPLs) as an undesired output. Our results show that the relationship between super-efficiency and stock market valuation is stronger than that between Return on Asset (ROA) and stock market performance, as measured by Tobin's Q. Notably, the Super-SBM-UND-VRS model yields novel results compared to other efficiency methods, such as the Stochastic Frontier Analysis (SFA) approach and traditional DEA models. Furthermore, our results suggest that bank evaluations benefit from decreased ownership concentration, whereas interest rate liberalization has the opposite effect.

# 1.Introduction

Evaluating the efficiency of commercial banks has long been a critical concern for regulators, bank managers, and investors. While traditional financial ratio analysis provides some insights, it falls short of revealing a comprehensive picture of financial institution performance. To address this limitation, various methods have been employed to study bank efficiency, including stochastic frontier analysis (SFA) and data envelopment analysis (DEA). SFA is a parametric approach that requires pre-specified functional forms and focuses on central tendencies, whereas DEA is a non-parametric method that assumes a linear relationship between inputs and outputs. Recent reviews have provided insightful discussions on DEA(Camanho, et al. (2023) )and SFA( Kumbhakar, et al. (2020)). However, a more nuanced approach is needed, one that accounts for side effects side effects such as non-performing loan ratios and leverage ratios. Moreover, a related question emerges: how do financial markets price bank stocks based on their efficiency performance? This paper aims to extend and integrate these two important questions, with a specific focus on the banking sector in China.

The debate surrounding the reliability of banking efficiency measurements remains ongoing. For example, Berger and Humphrey (1997) found that DEA efficiency scores present greater variability than SFA. Resti (1997) suggest that econometric and linear programming results do not differ dramatically, when based on the same data and conceptual framework. However, Bauer, et al. (1998) observed weak consistency in efficiency scores between nonparametric and parametric approaches, with nonparametric measures only weakly related to financial ratio performance measures. Beccalli, et al. (2006) suggest that changes in the prices of bank shares reflect percentage changes in cost efficiency, particularly those derived from DEA, compared to SFA efficiency and other control variables. More recently, Dong, et al. (2014) discovered moderate consistency between parametric and non-parametric frontier methods in efficiency scores rankings, whereas Silva, et al. (2017) identified a large divergence in the bank level estimates of SFA and DEA. This paper employs both approaches to compare these methods and investigate the relationship between efficiency and stock valuation.

The Chinese banking sector has been extensively studied, yielding a wealth of findings. For example, Berger, et al. (2009) found that minority foreign ownership of the Big Four banks is likely to significantly improve performance. Numerous studies have investigated China's bank efficiency, including those by Wang, et al. (2014), Hou, et al. (2014), Fungáčová, et al. (2019), Galán and Tan (2022). In light of the significant interest rate liberalization and mixed ownership reforms in China's banking sector in recent years, a reassessment is necessary to examine how these reforms have reshaped the industry.

Traditional data envelopment analysis (DEA) models, such as CCR (Charnes, et al. (1978)) and BCC(Banker, et al. (1984)), are hindered by several limitations, including difficulties in statistical testing, failure to identify avenues for efficiency improvement, and an inability to accurately reflect efficiencies above the production frontier. To address these limitations, this paper employs an non-directional and non-radial super-efficiency models based on Tone (2001) and Andersen and Petersen (1993), which offers the enhanced discrimination power and provides more nuanced insights into efficiency evaluation. Additionally, we utilize a parametric estimation procedure closely related to Battese and Coelli (1995) efficiency model. To the best of our knowledge, research examining the relationship between various measured efficiencies and stock market valuation is scarce, with studies focused on China's listed banks' valuation being virtually non-existent. This paper aims to illuminate the relationship between financial market valuation and efficiency analysis in this field.

In the second stage of our research, we seek to bridge the gap between super-efficiency scores derived from DEA models and SFA models, as well as market valuation of commercial banks. Specifically, we investigate whether investors can discern efficiency and its subsequent impact on market prices. We conduct a comparative analysis of the effects of various types of efficiency scores on bank stock performance, thereby identifying the most influential factors driving evaluation at the bank level.

The remainder of this paper is organized as follows: Section 2 provides a concise review of methodologies for evaluating bank efficiency, encompassing the SUPER-SBM-DEA-VRS model, other DEA models, and the SFA model. Section 3 applies these methodologies to assess the efficiencies of China's listed banks. Section 4 presents the empirical results of a panel regression analysis examining the relationship between bank efficiency and stock market valuation, as measured by Tobin's Q. Finally, Section 5 concludes based on our findings.

## 2. Research methods and Data Sources

2.1 DEA models

The first two DEA methods we employed is follow CCR (Charnes, et al. (1978)), BCC( Banker, et al. (1984), Färe, et al. (2013), Phan, et al. (2018), Proença, et al. (2023) ). The conventional variable-returns-to-scale(VRS) cost minimization model of cost efficiency y, also known as the input-oriented DEA model, is also utilized. Additionally, we employ two Slacks-based measures of efficiency under variable returns-to-scale assumption, which developed by Tone (2001). The key distinction between the SBM-VRS and SBM-UND-VRS models lies in their treatment of non-performing loans, which are either used as inputs or undesired outputs.

Regarding the implementation of the SUPER-SBM-VRS approach, we adopt the model outlined in Appendix 1, which belongs to the family of the Slacks-based DEA models.

Following Liu and Tone (2008)[2] input choices, this study selects tier one capital, interest expenses, and operating expense minus capital/credit loss provision along with capital/credit loss provision as inputs. In the production approach (PA), we evaluate the profit efficiency of commercial banks from a risk-return perspective, considering two outputs: net profit returned to the parent after consolidation, and non-performing loans as an undesirable by-product. Another approach, the intermediary approach (IA) assesses the operating efficiency of banks by considering deposits and loans as intermediate products. The intermediary approach (IA) use deposits and loans, and non-performing loans as three outputs[3].

2.2 Stochastic Frontier Approach

This study follows the methodology established by (Aigner, et al. (1977); Meeusen and van Den Broeck (1977); and developed by Battese and Coelli (1995); Eisenbeis, et al. (1999), Demerjian, et al. (2012); Sun, et al. (2013); Silva, et al. (2017), Bensalem and Ellouze (2019), Han, et al. (2024)). To ensure comparability with DEA models, we define the output and inputs consistently. In line with Berger and Mester (1997)[4] profit efficiency concept, we specify the inputs as the price of funds (interest expenses scaled by total deposits, w1) and the cost of loan (non-performing loan scaled by loans, w2), along with capital/credit loss provision and general and administrative expenses. Consequently, the input variables in the SFA model comprise two additional variables: deposits and loans.

$$\ln(\pi + \theta) = f(w, p, z, v) + \ln u_\pi + \ln \varepsilon_\pi \tag{1.1}$$

Where $\pi$ denotes net profits, *w1/w2* represents deposits and loans, *p1/p2 denotes* interest expenses scaled by total deposits and NPL ratio, and we incorporate capital/credit loss provision and general and administrative expenses in translog form into the model.

2.3 Data and Variable Selection

This study utilizes the annual report data of 42 listed banks in China as its primary data source, covering the period from 2006 to 2023 and yielding a total of 420 observed values. The sample consists of three categories of banks: 6 large state-owned banks (SOBs), 9 joint-stock banks (JSBs), and 27 urban commercial banks and rural commercial banks (typically smaller in size). A comprehensive list of the full names and abbreviations of the listed banks in China is provided in Appendix 2.

---

[2] Liu and Tone (2008) choose three inputs, namely: (1) interest expenses (IE); (2) credit costs (CC); and (3) general and administrative expenses (GAE), we add tier one capital as input.

[3] However, we find the link between IA approach and stock performance is relatively weak compare to the PA approach, for the concise need, the result of IA approach are available on request.

[4] According to Berger and Mester (1997), the profit efficiency is superior to the cost efficiency for evaluating the overall performance of the bank.

As shown in Table 1, we adopt five models to assess bank efficiency. The production approach (PA) also referred to as profit-oriented, measures efficiency related to the net profit excluding non-recurring gains and losses attributable to the parent company. Table 2 shows the descriptive statistics of the key variables,

Table 1: Input and output used in DEA and SFA models

| Models(PA) | X1 | X2 | X3 | X4 | X5 | Y1 | YB (Undesired Product) |
|---|---|---|---|---|---|---|---|
| SUPER-SBM-UND-VRS | Tier one capital | Interest expense | Operational expense minus capital/credit loss provision | Credit/capital Loss Provision | | Net profit | NPL |
| SBM-UND-VRS | Tier one capital | Interest expense | Operational expense minus capital/credit loss provision | Credit/capital Loss Provision | | Net profit | NPL |
| SBM-VRS | Tier one capital | Interest expense | Operational expense minus capital/credit loss provision | Credit/capital Loss Provision | NPL | Net profit | |
| Input-oriented BCC | Tier one capital | Interest expense | Operational expense minus capital/credit loss provision | Credit/capital Loss Provision | NPL | Net profit | |
| SFA | | Interest expense | Operational expense minus capital/credit loss provision | Credit/capital Loss Provision | NPL | Net profit | |

Table 2: Description statistics for key Variables

| | Details | N | Mean | SD | Min | p25 | p75 | Max |
|---|---|---|---|---|---|---|---|---|
| Tobinsq | Tobins'Q | 420 | 2.423 | 1.87 | 1.184 | 1.594 | 2.449 | 19.123 |
| T1 | Tier one capital | 420 | 4086.336 | 6498.382 | 64.74 | 357.365 | 4838.79 | 37765.898 |
| ie | Interest expense | 420 | 929.795 | 1207.965 | 10.59 | 120.271 | 1276.16 | 7500.26 |
| oec | Operational expense minus capital/credit loss provision | 420 | 509.726 | 722.645 | 7.01 | 40.72 | 556.04 | 3165.75 |
| ccloss | Credit/capital Loss Provision | 420 | 283.399 | 392.844 | 0.99 | 33.23 | 434.33 | 2026.68 |
| netprofit | Net profit | 420 | 498.616 | 757.897 | 6.24 | 42.708 | 542.55 | 3614.11 |
| npl | Non-performing loan | 420 | 409.929 | 656.118 | 1.314 | 32.864 | 535.985 | 3535.02 |
| size | Natural logarithm of total assets | 420 | 9.875 | 1.561 | 6.627 | 8.642 | 11.112 | 13.01 |
| deposits | Deposits | 420 | 37876.763 | 58910.457 | 509.32 | 3606.995 | 40406.299 | 335212 |
| loan | Loan | 420 | 29325.626 | 45263.473 | 306.29 | 2564.755 | 35737.85 | 253869 |

**Note**: The variable tobinsq is calculated as the book value of total assets minus the book value of common equity plus the market value of common equity, all scaled by the book value of assets. The control variables are defined as follows: firm size (SIZE), measured by the natural logarithm of total assets; and all other variables are in units of 100 million RMB yuan, excluding size and tobinsq.

## 3. Empirical results

3.1 Efficiencies of listed Banks and *Tobins'Q* (2006-2023)

Figures 1 and 2 illustrate the average efficiency scores of the Chinese banking system, as estimated by the Super-SBM-Und-VRS model and Stochastic Frontier Analysis (SFA), respectively. Our results reveal a decline in average DEA scores starting from 2017, coinciding with the initiation of interest rate liberalization reforms in China. Similarly, SFA scores exhibit a downward trend commencing from 2017, although with a smoother trajectory. Additionally, we find that DEA efficiency scores display greater variability compared to SFA scores, consistent with Berger and Humphrey (1997) observation. Figure 2 shows that the Super-SBM-Und-VRS model records the highest variation among the four DEA methods..

Figure 3 displays the average *TOBINSQ* of the Chinese banking system. We observe a dramatic rise and fall in *TOBINSQ* from 2006 to 2010, partly attributed to the 2008 global financial crisis. Given that the crisis may seriously bias our empirical results, we will separate the data from 2010 in the robustness test.

Figure 1：The average efficiency scores of the chinese listed bank(I)

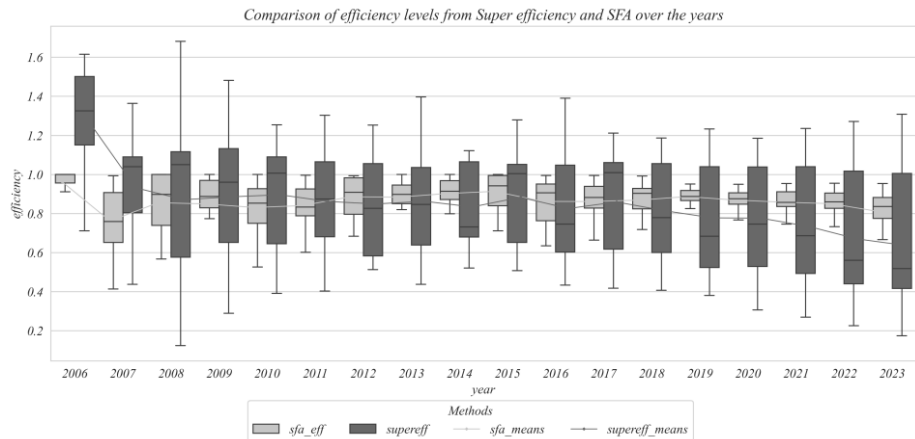

Figure 2：The average efficiency scores of the chinese listed bank(II)

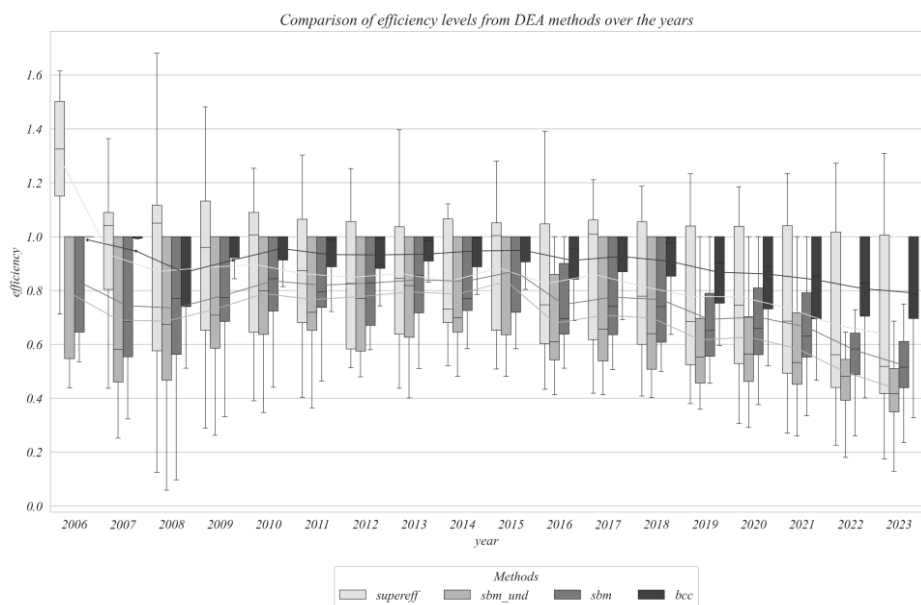

Figure 3: The average *Tobinsq* of the China listed bank over the years

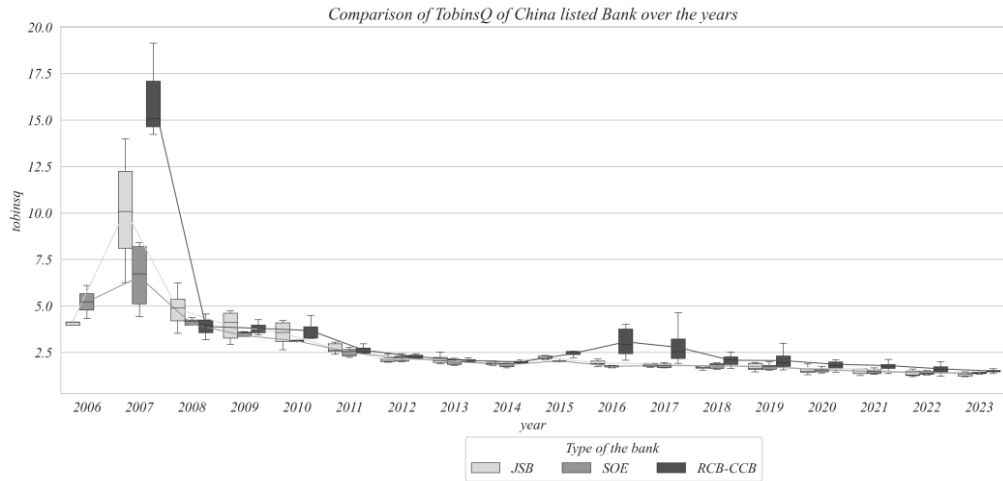

To investigate the similarity between individual efficiency scores generated by the two methodologies, we compute Spearman's rank correlation coefficients between the DEA economic efficiencies and the SFA efficiency. Table 3 presents the pairwise Spearman rank order correlation coefficients between the profit efficiency scores obtained from each method and Tobin's Q. Notably, we observe the highest correlations between super-efficiency and Tobin's Q as 0.235. Furthermore, our results show moderate positive rank order correlations between the different efficiency scores, all of which are significant at the 1% level. When comparing the parametric technique with the non-parametric techniques, our findings suggest that SFA and traditional DEA exhibit moderate consistency in their rankings, with a rank order correlation coefficient exceeding 0.444, consistent with Dong, et al. (2014)'s result(0.422). As expected, the four DEA methods record high correlations due to the identical input and output settings. However, we find relatively weak correlations between SFA efficiency and Tobin's Q.

Table 3: Spearman's rank order correlation by various models and Tobin's Q

|           | tobinsq   | sfa eff   | supereff | sbm und  | sbm      | bcc |
|-----------|-----------|-----------|----------|----------|----------|-----|
| tobinsq   | 1         |           |          |          |          |     |
| sfaeff    | -0.034    | 1         |          |          |          |     |
| supereff  | 0.235***  | 0.473***  | 1        |          |          |     |
| sbmund    | 0.212***  | 0.496***  | 0.761*** | 1        |          |     |
| sbm       | 0.200***  | 0.508***  | 0.771*** | 0.994*** | 1        |     |
| bcc       | 0.198***  | 0.453***  | 0.866*** | 0.732*** | 0.769*** | 1   |

**Note:** The variable tobinsq is calculated as the book value of total assets minus the book value of common equity plus the market value of common equity, all scaled by the book value of assets. sfa eff denotes SFA efficiency; supereff denotes super-efficiency calculated by Super-SBM-UND-VRS model; sbmund denotes DEA efficiency calculated by SBM-UND-VRS model, sbm denotes DEA efficiency calculated by SBM-VRS model, bcc denotes DEA efficiency calculated by BCC model, t statistics in parentheses *p < 0.1, ** p < 0.05, *** p < 0.01.

## 4. Efficiency and Stock Market Valuation

4.1 Does profit efficiency affect corporate value?

In the second stage, we investigate the relationship between efficiencies and stock performance by regressing Tobin's Q against efficiency estimates and selected performance measures. Consistent with Beccalli, et al. (2006) findings, we expect changes in efficiency are reflected in changes in stock prices and that stocks of cost efficient banks tend to outperform their inefficient counterparts. Following the approach of Baker, et al. (2003), , McLean, et al. (2012), Shaukat and Trojanowski (2018), Brahma, et al. (2021), Zareie, et al. (2024), we estimate *TOBINSQ* as the book value of total assets minus the book value of common equity plus the market value of common equity, all scaled by the book value of assets. We also use future (i.e. next year) *TOBINSQ* in our robustness tests. The estimated models are:

Model 1: Stock Performance and Bank Efficiency

$$TobinQ_{i,t} = \beta_0 + \beta_1 E_{i,t} + \varepsilon_{i,t} \tag{1.2}$$

Model 2: Stock Performance, Bank Efficiency, and Proxies for size, return, leverage ratio.

$$TobinQ_{i,t} = \beta_0 + \beta_1 E_{i,t} + \beta_2 FIRMCTRL_{i,t} + \varepsilon_{i,t} \tag{1.3}$$

Model 3: Stock Performance, Bank Efficiency, and Proxies for size, return, leverage ratio, and gdp growth and spread(one year loan prime rate minus government bond rate) .

$$TobinQ_{i,t} = \beta_0 + \beta_1 E_{i,t} + \beta_2 FIRMCTRL_{i,t} + \beta_3 MACROCTRL_{i,t} + \varepsilon_{i,t} \tag{1.4}$$

Our main variable of interest is the efficiencies, following D'Costa and Habib (2024) , We control for various firm-level characteristics, denoted by FIRMCTRL, which includes: firm size (*SIZE*), measured by the natural logarithm of total assets; leverage ratio (*LEVR*), measured by the ratio of total debt to total assets, proxying for leverage; firm growth (Growth), measured by the asset growth, representing firm expansion; NPL ratio, serving as proxy of bank risk; bank type(TYPE), set of dummy variables indicate the type of bank. Additionally, we control for other bank-level characteristics, including: Niiratio, measured by noninterest income scaled by total income; Tenclient, measured by the percentage of the ten largest loan clients, representing client concentration; Tenowner, measured by t the percentage of the ten largest stockholders, capturing ownership concentration. At the MACROCTRL level, we control for: RGDP, measured by GDP growth rate; and SPREAD, measured by the one-year loan prime rate minus the one-year China government bond rate, which reflects the impact of interest rate liberalization reform in China.

Based on the Variance Inflation Factor (VIF) multi-collinearity test, we select two efficiencies - SFA efficiency and super efficiency - from among the five efficiencies to mitigate potential collinearity issues.[5] We expect efficiencies, Roa, rgdp and spread to be positively related with *TobinsQ*; while Size and growth rate to be negatively related with *TobinsQ*. To minimize the influence of outliers, we winsorize all the continuous variables at the extreme 1% of their respective distributions.

Table 4: Efficiencies and Stock Performance- Baseline Regression

| Dep.var | (1) TobinsQ | (2) TobinsQ | (3) TobinsQ | (4) TobinsQ | (5) TobinsQ | (6) TobinsQ |
|---|---|---|---|---|---|---|
| sfa_eff | -3.09*** | -1.34 | -1.32 | 0.00 | 0.29 | 0.29 |
|  | (-3.47) | (-1.24) | (-1.27) | (0.00) | (0.44) | (0.44) |
| supereff | 1.91*** | 0.65*** | 0.58** | 0.89*** | 0.60** | 0.60** |
|  | (6.23) | (3.13) | (2.68) | (3.75) | (2.67) | (2.67) |
| size |  | -0.14*** | -0.04 | -0.13** | -0.52** | -0.52** |
|  |  | (-5.30) | (-0.73) | (-2.61) | (-2.49) | (-2.48) |
| roa |  | 129.07** | 135.02*** | -16.35 | -97.72*** | -97.72*** |
|  |  | (2.66) | (2.82) | (-0.47) | (-3.71) | (-3.70) |
| levr |  | 34.95*** | 32.38*** | 18.96*** | 14.46** | 14.46** |
|  |  | (5.93) | (6.68) | (4.72) | (2.66) | (2.65) |
| nplratio |  | 1.01*** | 1.02*** | 0.71*** | 0.71*** | 0.71*** |
|  |  | (11.20) | (11.34) | (8.06) | (8.06) | (8.04) |
| growth |  | 3.67*** | 3.60*** | 1.76*** | 0.53* | 0.53* |
|  |  | (8.55) | (7.77) | (3.73) | (1.72) | (1.72) |
| niiratio |  |  | -0.01 | 0.00 | 0.00 | 0.00 |
|  |  |  | (-1.16) | (0.33) | (0.56) | (0.56) |
| tenclient |  |  | 0.01 | 0.00 | 0.02* | 0.02* |
|  |  |  | (0.93) | (0.61) | (1.90) | (1.89) |
| tenown |  |  | -0.01 | -0.00 | -0.02** | -0.02** |
|  |  |  | (-1.49) | (-1.52) | (-2.10) | (-2.09) |
| gdpg |  |  |  | 6.58*** | 4.86*** | 4.86*** |
|  |  |  |  | (7.83) | (5.05) | (5.03) |
| spread |  |  |  | 44.99*** | 31.23*** | 31.23*** |
|  |  |  |  | (4.38) | (3.47) | (3.46) |
| _cons | 3.53*** | -31.22*** | -29.35*** | -16.99*** | -6.99 | -6.99 |
|  | (5.14) | (-5.24) | (-5.70) | (-4.38) | (-1.11) | (-1.10) |
| Type FE | No | No | No | Yes | No | Yes |
| Firm FE | No | No | No | No | Yes | Yes |
| _cons | Yes | Yes | Yes | No | No | No |
| $N$ | 420 | 377 | 377 | 377 | 376 | 376 |
| Adjusted $R^2$ | 0.08 | 0.59 | 0.60 | 0.72 | 0.78 | 0.78 |

**Note:** This table presents the descriptive statistics for the regression variables. Robust t-statistics are in brackets and are based on standard errors clustered by firm. * $p < 0.1$, ** $p < 0.05$, *** $p < 0.01$.

Our baseline regression results are presented in Table 4, where we report three specifications. In the first specification (Column 1), we present the regression results without time-variant firm characteristics. In the subsequent specifications (Columns 2-6), we augment our model with: In the other columns (column 2-column6), we augment our model with firm-level control variables (column 2 and column 3) and with macro control variables (column 4-column6).

Notably, across all columns, we find that the super efficiency coefficient is consistently positive and statistically significant at the 10% level. This suggests that investors perceive higher super-efficiency as a positive signal of bank value. In contrast, we observe that SFA efficiency does not exhibit a significant and positive coefficient in our models. With respect to control variables, our results show that:

Firm size exhibits a negative and statistically significant coefficient at the 1% level, confirming our expectation.

---

[5] Furthermore, we conduct additional regressions using alternative DEA-based efficiency metrics, including four alternative measures. The results indicate that the super-efficiency metric yields the strongest predictive power. Interested readers can access the underlying data upon request.

Return on assets (ROA) is positive and statistically significant at the 1% level in two models.

As expected, GDP growth and Spread exhibit positive and statistically significant coefficients (p < 0.01). Additionally, the growth of assets also obtains a positive and statistically significant coefficient (p < 0.1).

Furthermore, we find that ownership concentration (tenown) is associated with a negative coefficient (p < 0.05 in columns 5-6), which supports the notion that Mixed ownership reform can lead to higher valuation by reducing ownership concentration.

### 4.2 Robustness test

We first run robustness tests to validate the evidence of a positive association between efficiency and TOBINSQ and report the results in Table 5. While our empirical setting does not provide a natural experiment allowing us to attribute causality to our results, we attempt to limit the endogeneity bias by repeating our analysis after replacing *TOBINSQ* with future *TOBINSQ* as the dependent variable. Another reason we use future *TOBINSQ* is that all the annual report are reported in the next year which rationalized the relationship between future *TOBINSQ* and efficiency measurement.

Table 5: Efficiencies and Stock Performance: Robustness test I

| Dep.var | (1) $TobinsQ_{t+1}$ | (2) $TobinsQ_{t+1}$ | (3) $TobinsQ_{t+1}$ | (4) $TobinsQ_{t+1}$ | (5) $TobinsQ_{t+1}$ | (6) $TobinsQ_{t+1}$ |
|---|---|---|---|---|---|---|
| sfa_eff | -1.52*** | -0.70 | -0.63 | 0.15 | 0.29 | 0.29 |
|  | (-2.61) | (-0.93) | (-0.84) | (0.25) | (0.70) | (0.70) |
| supereff | 1.66*** | 0.47** | 0.45** | 0.65*** | 0.43** | 0.43** |
|  | (8.24) | (2.52) | (2.40) | (3.22) | (2.33) | (2.32) |
| size |  | -0.09*** | -0.07 | -0.18*** | -0.75*** | -0.75*** |
|  |  | (-3.72) | (-1.54) | (-3.03) | (-5.21) | (-5.20) |
| roa |  | 104.28** | 104.34** | 25.36 | -63.96** | -63.96** |
|  |  | (2.38) | (2.34) | (0.64) | (-2.05) | (-2.04) |
| levr |  | 25.35*** | 24.29*** | 17.79*** | 12.53*** | 12.53*** |
|  |  | (5.75) | (6.07) | (6.87) | (4.37) | (4.36) |
| nplratio |  | 0.51*** | 0.49*** | 0.34*** | 0.25** | 0.25** |
|  |  | (4.33) | (4.06) | (2.75) | (2.05) | (2.04) |
| growth |  | 2.37*** | 2.33*** | 1.34*** | 0.14 | 0.14 |
|  |  | (6.45) | (5.76) | (3.03) | (0.44) | (0.44) |
| niiratio |  |  | -0.00 | 0.01 | 0.01* | 0.01* |
|  |  |  | (-0.07) | (1.14) | (1.96) | (1.95) |
| tenclient |  |  | 0.01 | 0.00 | 0.01 | 0.01 |
|  |  |  | (1.12) | (0.75) | (1.54) | (1.53) |
| tenown |  |  | 0.00 | 0.00 | -0.01 | -0.01 |
|  |  |  | (0.04) | (0.16) | (-0.75) | (-0.75) |
| gdpg |  |  |  | 2.85*** | 0.76 | 0.76 |
|  |  |  |  | (6.77) | (1.64) | (1.64) |
| spread |  |  |  | 23.82*** | 8.52** | 8.52** |
|  |  |  |  | (5.43) | (2.42) | (2.41) |
| _cons | 2.12*** | -22.35*** | -21.71*** | -15.23*** | -2.88 | -2.88 |
|  | (4.66) | (-5.16) | (-5.41) | (-6.43) | (-0.90) | (-0.90) |
| Type FE | No | No | No | Yes | No | Yes |
| Firm FE | No | No | No | No | Yes | Yes |
| _cons | Yes | Yes | Yes | No | No | No |
| N | 378 | 335 | 335 | 335 | 331 | 331 |
| Adjusted $R^2$ | 0.15 | 0.61 | 0.62 | 0.69 | 0.82 | 0.82 |

**Note:** This table presents the descriptive statistics for the regression variables. Robust t-statistics are in brackets and are based on standard errors clustered by firm. * p < 0.1, ** p < 0.05, *** p < 0.01.

Table 5 presents the results of the next year's Tobin's Q regression, using the same variables as in the baseline test. We find consistent results for the efficiency measures. To further robustness, we also examine the regression in different time periods, with a focus on the post-2008 global financial crisis era. Additionally, we employ an alternative model specification, where the dependent variable is the change in Tobin's Q, denoted as :

$$diff.TobinQ_{i,t} = \beta_0 + \beta_1 diffE_{i,t} + \beta_2 FIRMCTRL_{i,t} + \varepsilon_{i,t} \quad (1.5)$$

Table 6: Efficiencies and Stock Performance: Robustness test II

| Dep.var | (1) diff.TobinsQ | (2) diff.TobinsQ | (3) diff.TobinsQ | (4) diff.TobinsQ | (5) diff.TobinsQ | (6) diff.TobinsQ |
|---|---|---|---|---|---|---|
| Diff supereff | 0.25* | 0.38** | 0.49*** | | | |
|  | (1.74) | (2.50) | (3.73) | | | |
| Size |  | 0.04*** | 0.39*** |  | 0.04*** | 0.39*** |
|  |  | (4.89) | (6.76) |  | (5.11) | (6.59) |
| Diff roe |  | -0.63** | -0.53** |  | -0.55* | -0.47* |
|  |  | (-2.36) | (-2.06) |  | (-1.90) | (-1.74) |
| Diff nplratio |  | 0.17* | 0.36*** |  | 0.15 | 0.33*** |
|  |  | (1.82) | (3.23) |  | (1.35) | (2.87) |
| growth |  | -0.45 | -0.08 |  | -0.47 | -0.09 |
|  |  | (-1.20) | (-0.27) |  | (-1.17) | (-0.26) |
| Diff sfa_eff |  |  |  | -0.02 | 0.12 | 0.39 |
|  |  |  |  | (-0.08) | (0.40) | (1.36) |
| _cons | -0.19*** | -0.58*** | -4.16*** | -0.19*** | -0.60*** | -4.15*** |
|  | (-10.67) | (-5.61) | (-7.05) | (-11.10) | (-5.64) | (-6.87) |
| Firm FE | No | No | Yes | No | No | Yes |
| _cons | Yes | Yes | No | Yes | Yes | No |
| N | 329 | 327 | 326 | 329 | 327 | 326 |
| Adjusted $R^2$ | 0.01 | 0.15 | 0.25 | -0.00 | 0.13 | 0.22 |

Note: This table presents the descriptive statistics for the regression variables. Robust t-statistics are in brackets and are based on standard errors clustered by firm. * $p < 0.1$, ** $p < 0.05$, *** $p < 0.01$.

Table 6 presents the results of the regression on the first-difference of Tobin's Q, employing a reduced sample of firms during the 2011-2023 period as a robustness check. Our findings, based on both SFA and DEA efficiency measurements, are consistent with Beccalli, et al. (2006)'s conclusion: specifically, the SFA efficiency estimates are not reflected in the market as being equally important when compared to the super-efficiency estimates.

## 5. Conclusions

This study contributes to the intersection of capital market research and bank efficiency literature by examining the relationship between bank profit efficiency and stock market evaluation. Our results suggest that changes in bank valuation are significantly associated with percentage changes in profit efficiency, particularly those derived from DEA models. Notably, our empirical findings support the notion that super-efficiency is the best proxy among all DEA models, given the superior discriminant power (variance) of super-efficiency in all models. In contrast, the relationship between SFA efficiency estimates and stock market evaluation is less clear-cut. Moreover, our analysis reveals that other variables such as size, riskiness, and profitability have mixed effects on stock evaluation. Additionally, we find that major banking reforms in China, including mixed ownership reform and interest rate liberalization, have opposing effects on stock market evaluation, with positive and negative impacts, respectively.

Appendix 1: Super-SBM-UND-VRS based DEA model

The following is a prototype of the SBM model with an undesirable output[6]:

$$\rho^* = \min \frac{1 - \frac{1}{m}\sum_{i=1}^{m}\frac{s_i^-}{x_{i0}}}{1 + \frac{1}{s_1+s_2}\left(\sum_{r=1}^{s_1}\frac{s_r^g}{y_{r0}} + \sum_{r=1}^{s_2}\frac{s_r^b}{y_{r0}}\right)}$$

$$S.T \begin{cases} x_0 = X\lambda + s^- \\ y_0^g = Y^g\lambda - s^g \\ y_0^b = Y^b\lambda + s^b \\ s^-, s^g, s^b, \lambda \geqslant 0 \end{cases} \quad (1)$$

Considering that there are m inputs (x), $s_1$ kind of expected outputs ($y^g$) and $s_2$ kinds of undesired outputs ($y^b$), output y is ($y^g$, $y^b$), which represents expected outputs and undesired outputs respectively. $s^- \epsilon R^m$ and $s^b \epsilon R^{s_2}$ represents excess of input and undesirable output, and $s^g \epsilon R^{s_1}$ as shortage of output. The current decision unit is efficient if and only $\rho$ =1, and $s^-$, $s^g$, $s^b$ are zeros. When the three relaxation conditions are not all zeros, the decision-making unit lacks efficiency. The form of the improved super efficiency SBM model is:

$$\rho^* = \min \frac{\frac{1}{m}\sum_{i=1}^{m}\frac{\bar{x}}{x_{i0}}}{\frac{1}{s_1+s_2}\left(\sum_{r=1}^{s_1}\frac{\bar{y}_r^g}{y_{r0}} + \sum_{r=1}^{s_2}\frac{\bar{y}_r^b}{y_{r0}}\right)}$$

$$S.T \begin{cases} \bar{x} \geqslant \sum_{j=1, \neq 0}^{n} \lambda_j x_j \\ \bar{y}_r^b \geqslant \sum_{j=1, \neq 0}^{n} \lambda_j y_j^b \\ \bar{y}_r^g \leqslant \sum_{j=1, \neq 0}^{n} \lambda_j y_j^g \\ \bar{x} \geqslant x_0, \bar{y}_r^b \geqslant y_0^b \text{ and } \bar{y}_r^g \leqslant y_0^g \\ \sum_{j=1, \neq 0}^{n} \lambda_j = 1 \ (VRS \ condition) \\ y \geqslant 0, \lambda \geqslant 0 \end{cases} \quad (2)$$

In order to solve the problem that the method of Super-SBM in some cases has no feasible solution, this paper refers to the method of Fang, et al. (2013) for two-stage solution. The super-efficiency are evaluated as:

---

[6] Page 318 of Cooper, et al. (2007)

$$SE = \begin{cases} 1 - \dfrac{1}{m}\sum_1^m \dfrac{s_i^-}{x_i} - \dfrac{1}{k}\sum_1^m \dfrac{s_r^b}{y_i}; & if\ SE < 1 \\ 1 + \dfrac{1}{m}\sum_1^m \dfrac{s_i^-}{x_i} + \dfrac{1}{k}\sum_1^m \dfrac{s_r^b}{y_i}; & if\ SE > 1 \end{cases} \qquad (3)$$

Appendix 2: English name and Chinese name of Listed Banks in China（2006-2023）

| Code(SHSE) | Dmu | Chinese Name | English Name | Abbr |
|---|---|---|---|---|
| 1 | 1 | 平安银行 | Ping An Bank Co., Ltd. | PABC |
| 1227 | 2 | 兰州银行 | Bank Of Lanzhou Co.,Ltd. | BLZC |
| 2142 | 3 | 宁波银行 | Bank Of Ningbo Co.,Ltd. | BNC |
| 2807 | 4 | 江阴银行 | Jiangsu Jiangyin Rural Commercial Bank Co.,Ltd. | JJRCB |
| 2839 | 5 | 张家港农商行 | Jiangsu Zhangjiagang Rural Commercial Bank Co., Ltd | JZRCB |
| 2936 | 6 | 郑州银行 | BANK OF ZHENGZHOU CO. ,LTD. | BZZC |
| 2948 | 7 | 青岛银行 | BANK OF QINGDAO CO., LTD. | BQDC |
| 2958 | 8 | 青农商行 | Qingdao Rural Commercial Bank Corporation | QDRCB |
| 2966 | 9 | 苏州银行 | Bank Of Suzhou Co.,Ltd | BSZ |
| 600000 | 10 | 浦发银行 | Shanghai Pudong Development Bank Co.,Ltd. | SPDB |
| 600015 | 11 | 华夏银行 | Hua Xia Bank Co.,Limited | HB |
| 600016 | 12 | 民生银行 | China Minsheng Banking Corp., Ltd. | CMSB |
| 600036 | 13 | 招商银行 | China Merchants Bank Co., Ltd. | CMB |
| 600908 | 14 | 无锡银行 | Wuxi Rural Commercial Bank Co., Ltd. | WXRCB |
| 600919 | 15 | 江苏银行 | Bank Of Jiangsu Co.,Ltd. | BOJS |
| 600926 | 16 | 杭州银行 | Bank Of Hangzhou Co.,Ltd. | BOHZ |
| 600928 | 17 | 西安银行 | BANK OF XI'AN CO., LTD. | BOXA |
| 601009 | 18 | 南京银行 | Bank Of Nanjing Co.,Ltd. | BONJ |
| 601077 | 19 | 重庆农商行 | Chongqing Rural Commercial Bank Co., Ltd. | CQRCB |
| 601128 | 20 | 常熟银行 | Jiangsu Changshu Rural Commercial Bank Co., Ltd. | CSRCB |
| 601166 | 21 | 兴业银行 | Industrial Bank Co.,Ltd. | IBC |
| 601169 | 22 | 北京银行 | Bank Of Beijing Co.,Ltd. | BOB |
| 601187 | 23 | 厦门银行 | Xiamen Bank Co.,Ltd. | XMB |
| 601229 | 24 | 上海银行 | Bank of Shanghai Co., Ltd. | BOSH |
| 601288 | 25 | 农业银行 | Agricultural Bank Of China Limited | ABC |
| 601328 | 26 | 交通银行 | Bank of Communications Co.,Ltd. | BC |
| 601398 | 27 | 工商银行 | Industrial And Commercial Bank Of China Limited | ICB |
| 601528 | 28 | 瑞丰农商 | Zhejiang Shaoxing RuiFeng Rural Commercial Bank Co.,Ltd | RFRCB |
| 601577 | 29 | 长沙银行 | BANK OF CHANGSHA CO., LTD | BCS |
| 601658 | 30 | 邮储银行 | POSTAL SAVINGS BANK OF CHINA CO., LTD. | PSB |
| 601665 | 31 | 齐鲁银行 | QILU BANK CO., LTD. | QLB |
| 601818 | 32 | 光大银行 | China Everbright Bank Company Limited Co., Ltd | CEB |
| 601825 | 33 | 上海农商 | Shanghai Rural Commercial Bank Co., Ltd. | SHRCB |
| 601838 | 34 | 成都银行 | Bank Of Chengdu Co.,Ltd. | BCD |
| 601860 | 35 | 紫金银行 | Jiangsu Zijin Rural Commercial Bank Co.,Ltd. | ZJRCB |
| 601916 | 36 | 浙商银行 | CHINA ZHESHANG BANK CO., LTD. | CZSB |
| 601939 | 37 | 建设银行 | China Construction Bank Corporation | CCB |
| 601963 | 38 | 重庆银行 | BANK OF CHONGQING CO., LTD. | BOCC |
| 601988 | 39 | 中国银行 | Bank Of China Limited | BOC |
| 601997 | 40 | 贵阳银行 | Bank Of Guiyang Co.,Ltd. | BOGY |
| 601998 | 41 | 中信银行 | CHINA CITIC BANK CORPORATION LIMITED | CITIC |
| 603323 | 42 | 苏州农商 | Jiangsu Suzhou Rural Commercial Bank Co.,Ltd. | SZRCB |